# Analyzing cases of significant nondynamic correlation with DFT using the Atomic Populations of Effectively Localized Electrons

C. Lewis[1], E. Proynov[1,2], Jianguo Yu[1] and J. Kong[1,2]*

[1] Department of Chemistry, Middle Tennessee State University, Murfreesboro, Tennessee, 37132 USA
[2] Center for Computational Sciences, Middle Tennessee State University, Murfreesboro, Tennessee, 37132 USA

*Correspondence: jing.kong@mtsu.edu

**Abstract**
Multireference effects are associated with degeneracies and near-degeneracies of the ground state and are critical to a variety of systems. Most approximate functionals of density functional theory (DFT) fail to properly describe these effects. A number of diagnostics have been proposed that allow to estimate the reliability of a given single-reference solution in this respect. Some of these diagnostics however lack size-consistency, while remaining computationally expensive. In this work we use the DFT method of determining atomic populations of effectively localized electrons (APELE) as a novel diagnostic in this vein. It is compared with several existing diagnostics of nondynamic correlation on select exemplary systems. We show that the APELE method is on average in good agreement with the existing diagnostics, while being both size-consistent and less costly. It becomes particularly informative in cases involving bond stretching or bond breaking. The APELE method is applied next to organic diradicals like the bis-acridine dimer and the p-quinodimethane molecule which possess unusually high nonlinear optical response, and to the reaction of ethylene addition to Ni dithiolene, where our results shed some more light on how the oxidation state of the Ni center may change when going from the initial reactant to the product.

## 1. Introduction

In a typical theoretical study, one would start with some popular functional of density functional theory (DFT) or with a low level wavefunction-based correlation method and eventually would advance to the coupled-cluster singles and doubles with perturbative triples (CCSD(T)) if the results are in doubt. These methods assume the dominance of a single reference configuration (single determinant- based methods). Single determinant methods are unreliable for systems with strong multireference character in which more than one electronic configuration plays a significant role. The electron correlation due to multireference effects is called nondynamic correlation (NDC), and strong correlation when the NDC largely dominates. In multicenter systems NDC tends to decouple a bond electron pair to two single unpaired electrons maximally separated from each other. This effect is also known as 'left-right' correlation [1-3]. The "simplest" example is the dissociating hydrogen molecule. At equilibrium, its two electrons with opposite spins are paired in a single bond pair. Upon dissociation, the electrons become more and more "effectively unpaired" in the sense of virtual unpairing and localization on the H centers. This is due to the increased left-right correlation upon dissociation, which provides energy gain by avoiding the two electrons to be close by. A proper description of this dissociation should take such a transformation into account without breaking the spin symmetry of the state.

Mainstream contemporary DFT functionals are known to poorly treat nondynamic/strong correlation, while some recent developments have shown great promise [4-9]. Instead, wavefunction-based methods like CASSCF (complete active space self-consistent field), MRCI (multi-reference configuration interaction), and their variations are usually employed. These methods are computationally expensive and can only be afforded to problems of relatively small size. It is desirable to be able to estimate *a priori*, the degree of multireference character of a given state, in order to estimate the reliability of a single-determinant method. A few single-determinant based descriptors (diagnostics) have been proposed for this purpose. The most commonly used are based on CCSD(T), such as the $T_1$ [10], $D_1$ [11], and %TAE[12] indexes. These methods estimate the strength of NDC in a system averaged way, averaging it per electron or per bond. However, NDC is often localized in a specific subdomain of the system, like on a transition-metal center in a complex or solid for example. In such cases, the NDC should remain quantitatively about the same upon increasing the size by adding larger ligands. Averaging the NDC over all the electrons and/or bonds is not a size-consistent approach since some of these may not be directly involved in the multireference process. This leads to an artificial dilution of the NDC estimate.

Alternative methods of treating such effects in a local fashion are the effectively-unpaired ('localized', 'odd') electron models of Yamaguchi [13], Bochicchio [14], Davidson [15], and Head-Gordon [16]. Upon stretching the bond between two nuclei, the bonding electrons become partially decoupled due to the increased left-right NDC and tend to virtually localize as unpaired electrons on the nuclei. These electrons are referred to as 'effectively localized electrons' (ELE) here. One example of ELE are the electrons with radical character in diradicals. The above methods include, in principle, both nondynamic and dynamic correlation, but localized radical density emerges mainly due to NDC. These methods require the correlated first-order reduced density matrix (RDM-1) which is not directly accessible from the single determinant Kohn-Sham (KS) or Hartree-Fock (HF) solution. As an alternative, Nakano et al [17] have proposed an index of diradical character ($y$) based on the HOMO-LUMO gap in natural-orbital (NO) representation of the spin-projected unrestricted SCF solution.

We have developed a DFT method of describing the formation of radical-like density [18]. It leads to a compact expression of the atomic population of effectively localized electrons (APELE) within a single-determinant KS DFT approach. The method retains the computational efficiency of a single-determinant KS solution while incorporating NDC in the evaluation of APELE. Calculations do not require spin-symmetry breaking, thus avoiding the side effects of spin contamination. By design, the APELE values reflect the strength of NDC locally which leads to size-consistent comparisons among systems of different sizes. A typical example is a singlet diradical, where two electrons with opposite spin, that formally would be paired in a bond pair, are in fact localized, each on a separate center. It requires two-reference configurations or more in a correlated wavefunction treatment of such a situation without spin-symmetry breaking. Larger APELE values of the radical atoms reflect stronger NDC acting locally among the radicalized electrons.

Indeed, strong electron correlation manifests most notably in singlet diradical molecules. Much attention has been devoted recently to a particular class of organic diradicaloids that have unusually high nonlinear optical (NLO) response [19]. Molecular design of efficient NLO systems is a subject of intense research due to their importance for photonics and optoelectronics such as optical switches, 3D microfabrics, photodynamic therapy and others. Recent studies have shown that the degree of diradicaloid character in systems like diphenalenyls, π-electron systems involving imidazole rings and other compounds that involve quinoid and benzenoid resonance structures like p-quinodimethane (PQM), is a key factor for the large second order hyperpolarizability ($\gamma$) of these systems [17, 20-23]. Similar features have been found also in some graphene nanoflakes [24, 25]. A thorough theoretical analysis has revealed that this large NLO enhancement is due to the peculiar electronic structure of the diradicaloids.

## 2. The APELE method within the Kohn-Sham DFT

The formation of open-shell singlet systems is driven by strong left-right NDC that tends to partly decouple a covalent bond pair and form radical-like ELE. This formation has been related to the non-idempotency of the correlated RDM-1 [13, 26]. At single-determinant SCF level (HF or KS) RDM-1 remains idempotent (integer orbital occupation numbers of 1 or 0) and an ELE formation is not feasible at that level. We have proposed a way to remedy this situation [18]. The (non) idempotency condition of the RDM-1 $\gamma(\mathbf{r};\mathbf{r}`)$ reads:

$$D_\sigma(\mathbf{r}) = \rho_\sigma(\mathbf{r}) - \int \gamma_\sigma(\mathbf{r};\mathbf{r}`)\gamma_\sigma(\mathbf{r}`;\mathbf{r})d\mathbf{r}` \geq 0 \qquad (1)$$

where $\rho_\sigma$ is the electron density of spin $\sigma$, $D_\sigma$ is the density of ELE per spin $\sigma$. The total ELE density $D_u$ and the gross mean number of ELE $\bar{N}_u$ are then given by:

$$D_u(\mathbf{r}) = 2\sum_\sigma D_\sigma(\mathbf{r}); \qquad \bar{N}_u = \int D_u(\mathbf{r})d\mathbf{r} \qquad (2)$$

At the single-determinant level (KS or HF) RDM-1 is idempotent meaning $D_\sigma \equiv 0$ everywhere as a limiting case of Eq. (1). We generalize the integrand of Eq.(1) by introducing a model effective exchange-hole that is otherwise compatible with the single-determinant restricted KS scheme but having a specific form [18]:

$$\gamma_\sigma(\mathbf{r};\mathbf{r}')\gamma_\sigma(\mathbf{r}';\mathbf{r}) \approx \rho_\sigma(\mathbf{r})\bar{h}_{X\sigma}^{eff}(\mathbf{r},\mathbf{r}') \qquad (3)$$

such that:

$$\int \bar{h}_{X\sigma}^{eff}(\mathbf{r},\mathbf{r}')d\mathbf{r}' \leq 1 \qquad (4)$$

The existence of such a form was justified using an approximate cumulant expansion of the two-body density matrix [18]. In the APELE model the effective exchange hole in Eq.(3) is represented by the relaxed exchange hole of Becke-Roussel [27] introduced in the B05 functional formalism [3, 28], $h_{X\sigma}^{B05}$, which estimates the NDC locally by its relaxed normalization $\bar{N}_{X\sigma}^{eff}$ (similar to Eq.(4)):

$$\int h_{X\sigma}^{B05}(\mathbf{r},\mathbf{r}')d\mathbf{r}' = \bar{N}_{X\sigma}^{eff}(\mathbf{r}) \leq 1 \qquad (5)$$

The smaller the relaxed normalization $\bar{N}_{X\sigma}^{eff}$, the stronger the NDC locally at a given reference point r. Using this relaxed exchange hole normalization in Eqs. (3),(1), the ELE density takes the form:

$$D_u(\mathbf{r}) = 2\sum_\sigma \rho_\sigma(\mathbf{r})\left(1 - \bar{N}_{X\sigma}^{eff}(\mathbf{r})\right) \qquad (6)$$

The smaller the relaxed normalization $\bar{N}_{X\sigma}^{eff}$, the larger the ELE density.

To obtain the APELE of a given atom $A$ ($F_r(A)$) we partition the molecular space into atomic regions using the grid-based atomic partitioning ($\Omega_A$) of Becke's grid integration scheme [29]

$$F_r(A) = \int_{\Omega_A} D_u(\mathbf{r})d\mathbf{r}, \qquad \sum_A F_r(A) = \bar{N}_u \qquad (7)$$

The values of individual APELE, $F_r(A)$, and the gross number of ELE, $\bar{N}_u$, can be used as alternative diagnostics measuring the strength of NDC locally ($F_r(A)$) and globally ($\bar{N}_u$) per Eq.(7). The APELE analysis was applied previously to elucidate more fully the bonding nature of $Cr_2$, $C_2$ and $(NO)_2$ dimer [18]. It was found that the bond in $C_2$ is unlikely to have diradical character in its ground state, while the singlet ground state of the (NO)$_2$ dimer does have a diradical character. The APELE method correctly estimated the magnetic moment localized on the Cr atoms in Cr$_2$ and predicted that the bond in Cr$_2$ may have a quad-radical nature [18].

## 3. Computational details

The APELE based NDC indexes were calculated with the KP16/B13 functional [5] which take into account NDC via real-space corrections to the exchange hole. This allows open-shell singlets to be described with restricted single-determinant KS (RKS) solution even at large bond stretches. This way we avoid the painful symmetry breaking procedure that is not always feasible.

Psi4, an open-source quantum chemistry software package [30], was used to optimize the structures at B3LYP/cc-pVQZ level using very tight convergence criteria. Psi4 was also used to evaluate the CCSD(T) diagnostics $T_1$, $D_1$, and %TAE[(T)], as well as, the $A_\lambda$ diagnostic for M06, M06-2X, and M06-L [30]. Instances in which the Psi4 package failed to deliver accurate results were circumvented using the Gaussian '09 software package [31]. The APELE results were obtained with the KP16/B13 functional using cc-pVTZ and G3LARGE basis sets [32] using the xTron code, an in-house computational program that allows the efficient computation of the HF exchange energy density [33-35]. It allows users to run various datasets with single commands in a cluster environment.

## 4. Diagnostics of Nondynamic Correlation.

Wavefunction-based methods that include both dynamic and nondynamic correlation are expensive and quickly out-scale the current computational resources. This motivated the need for diagnostic tools to determine whether NDC must be added to a given single-reference solution. There have been a number of techniques created for this purpose, however, they often involve the use of computationally complex methods themselves or require the help of other techniques. Here we consider a few of the more commonly used diagnostics together with the proposed APELE based diagnostics.

### 4.1. Summary of Current Diagnostics.

#### 4.1.1. Methods based on unrestricted SCF

Nakano et al [17] have proposed an index of diradical character ($y$) based on the HOMO-LUMO gap in natural-orbital (NO) representation of the spin-projected unrestricted SCF (HF or Kohn-Sham) solution.

$$y = 1 - \frac{2T}{1+T^2} , \qquad (8)$$

where $T$ is the orbital overlap between the HOMO and LUMO NOs. This analysis relies heavily on spin symmetry breaking, which brings large spin contamination. Moreover, a small HOMO-LUMO gap does not necessarily correspond to a large NDC, e.g. in a metallic system. Another problem is that systems with significant NDC do not always possess a spin-symmetry broken stable solution with energy lowering.

#### 4.1.2. CI based

The configuration interaction (CI) expansion of the wave function contains the expansion coefficients for each determinantal configuration involved. A system with a large amount of NDC would have a relatively small contribution to the total CI wave function from the main ground-state determinant. This diagnostic is represented by the leading coefficient ($C_0$), from the CI wave function expansion. It is quite informative, but it has an inherent bias tending to give higher $C_0$ values when the method uses single-determinant HF orbitals [36]. Complete Active Space SCF (CASSCF) calculations are used to remedy these shortcomings. It is generally accepted that a system with a $C_0 < 0.95$ or $C_0^2 < 0.90$ has multireference character and a multireference method should be used. The CASSCF solution however takes a considerable amount of time for all but small systems.

### 4.1.3. Coupled cluster based

The coupled cluster method employs *a priori* measures based on the $\vec{t}_1$ amplitudes and/or uses %TAE[(T)] (atomization energy due to triple excitations) as indicators of near-degeneracy. The Frobenius norm of the $\vec{t}_1$ amplitudes taken as a base and divided by the square root of the number of correlated electrons define the $T_1$ diagnostic [10]:

$$T_1 = \frac{\|\vec{t}_1\|}{\sqrt{n}} \quad (9)$$

The 'number of correlated electrons' $n$ here implies only the valence electrons but not the core electrons. A few thresholds for $T_1$ have been suggested, which are largely based on the types of systems studied [10, 36] It is generally accepted that $T_1 > 0.02$ for organic molecules, $T_1 > 0.05$ for 3d transition metals, and $T_1 > 0.045$ for 4d transition metals indicates severe NDC. [36, 37]

The $D_1$ diagnostic is based on the matrix norm of the $\vec{t}_1$ amplitudes. Instead of using the sum of the amplitudes, it uses the largest one: [11]

$$D_1(\text{CCSD}) = \|T\|_{(2)} \quad (10)$$

$D_1$ was developed because $T_1$ alone does not describe well certain systems that possess high degree of NDC. A NDC threshold of $D_1 > 0.15$ was suggested for 3d transition metals and $D_1 \geq 0.12$ for 4d transition metals. Taking these two diagnostics together, one may also use the ratio $T_1/D_1$ which sheds light on the degree of homogeneity of the system [38].

Another diagnostic rooted in the coupled cluster approach is %TAE[(T)]. It uses the atomization energy from CCSD(T) compared with that from CCSD. This gives the percentage of atomization energy coming from the triple excitations alone:

$$\%\text{TAE}[(T)] = \frac{\text{TAE}(\text{CCSD}(T)) - \text{TAE}(\text{CCSD})}{\text{TAE}(\text{CCSD}(T))} * 100 \quad (11)$$

The %TAE[(T)] index is an energy-based diagnostic. It has been shown to correlate well with the more comprehensive estimates based on higher levels of the theory, such as %*TAE*[($T_4 + T_5$)] **[36, 39]** . %TAE[(T)] indicates mild, moderate, or severe NDC at less than 5%, greater than 5% but less than 10%, and greater than 10%, respectively **[36]**. This classification remains consistent across the different types of molecules.

### 4.1.4 DFT based diagnostics

There are a few methods developed for gauging NDC within DFT that are primarily energy based. The $B_1$ method developed by Truhlar et al. has been shown to correlate well with the $T_1$ diagnostic for systems such as BeO and MgO [40]. Another diagnostic rooted in DFT that displays similar qualities is the $A_\lambda$ diagnostic [39]. It uses the ratio of the atomization energy from a hybrid functional vs the atomization energy with 100% of HF exchange alone:

$$A_\lambda = \frac{\left(1 - \frac{\text{TAE}[(X_\lambda C)]}{\text{TAE}[(X_{100})]}\right)}{\lambda} \quad (12)$$

The fraction of model DFT exchange in the hybrid functional acts to some extent as a partial substitute for the left-right correlation. The fraction of HF exchange $\lambda$ in the denominator normalizes this diagnostic. For $A_\lambda$ < 0.10 the system is primarily dominated by dynamic correlation, $A_\lambda \approx 0.15$ indicates that it is mildly ND correlated, $A_\lambda \approx 0.30$ means that it is moderately ND correlated and for $A_\lambda > 0.50$ the system should be strongly ND correlated. It has been shown that $A_\lambda$ correlate well with %TAE[(T$_4$ + T$_5$)] when certain functionals are used. A problem with $A_\lambda$ is that the results sensibly depend on the approximate functionals used.

### 4.2. Diagnostic based on the atomic population of effectively localized electrons (APELE).

The general idea of the APELE method was given in Section 2. Integrating the APELE over the whole space gives the gross number of ELE $\bar{N}_u$ in the system. Given the specific set-up of our model, this gross ELE number reflects the overall strength of NDC in the system and can be used as a complementary NDC diagnostic. Indexing the NDC based on the APELE analysis ($F_r(A)$ and $\bar{N}_u$) has some distinct advantages. Above all, the APELE is a local characteristic and does not include averaging over the whole system like in the case of $T_1$ or %TAE indexes. Another advantage is the computational efficiency since our method is based on single-determinant KS-DFT. Furthermore, the method directly accounts for the NDC by design, so that the APELE values directly reflect the strength of NDC locally ($F_r(A)$) or globally ($\bar{N}_u$).

#### 4.2.1 Linear alkane chains with a stretched central C-C bond distance

An example illustrating the size-consistency issue with the CCSD(T)-based indexes is a set of linear alkanes with a stretched carbon-carbon bond in the middle. We present our APELE results for ethane (C2H6), butane, (C4H10), hexane (C6H14) and octane (C8H18), in Fig.1 and Tables 1-4. The tables also contain the values of several other existing diagnostics calculated as explained in the computational details. The molecular geometries are first optimized with B3LYP/ cc-pVQZ. Then, the middle C-C bond is stretched out, starting from the equilibrium, up to 3.5˚A. The stretching induces strong left-right ND correlation mainly localized in the middle C-C bond of each molecule. Figure 1 show the dependence of APELE on one of the two central carbon atoms in each of these four alkanes with increasing the C-C distance. It is consistent with one's physical intuition that the NDC should increase when the bond is progressively stretched. This trend holds steady, and the values only slightly vary from ethane to octane, showing that the APELE index is size-consistent and reflects the increase in left-right correlation locally, mainly at the middle C-C bond.

The coupled-cluster-based diagnostics show the proper trend as the central C-C bond distance increases in each of the molecules. However, as we move from smaller to larger molecular systems, the $T_1$, $D_1$, and %TAE[(T)] values tend to decline when comparing them at the same C-C distance. As the size of the molecule increases, the values of these diagnostics become more diluted, which is a sign of size-inconsistency. A similar trend with the coupled-cluster-based diagnostics was reported in the literature as well [38].

**Table 1**: **Ethane**: Left-Right Correlation Analysis. The APELE values are per C atom of the central C-C bond.

| Diagnostic | 1.5Å | 2.0Å | 2.5Å | 3.0Å | 3.5Å |
|---|---|---|---|---|---|
| APELE | 0.1905 | 0.2262 | 0.2894 | 0.3754 | 0.4567 |
| $T_1$ | 0.0036 | 0.0088 | 0.0185 | 0.0340 | 0.0312 |
| $D_1$ | 0.0075 | 0.0324 | 0.0629 | 0.1304 | 0.1129 |
| %TAE[(T)] | -0.0501 | 0.1946 | 1.3395 | 2.9680 | 4.7210 |
| $A_\lambda$[M06] | -0.0088 | -0.0086 | 0.0021 | 0.0174 | 0.0297 |
| $A_\lambda$[M06-2X] | -0.0034 | -0.0035 | 0.0090 | 0.0289 | 0.0493 |
| $A_\lambda$[M06-HF] | -0.0059 | -0.0077 | 0.0019 | 0.0204 | 0.0407 |

**Table 2**: **Butane**: Left-Right Correlation Analysis. The APELE values are per C atom of the central C-C bond.

| Diagnostic | 1.5Å | 2.0Å | 2.5Å | 3.0Å | 3.5Å |
|---|---|---|---|---|---|
| APELE | 0.1968 | 0.2374 | 0.3061 | 0.3912 | 0.4679 |
| $T_1$ | 0.0041 | 0.0074 | 0.0156 | 0.0350 | 0.0437 |
| $D_1$ | 0.0104 | 0.0345 | 0.0653 | 0.1930 | 0.2074 |
| %TAE[(T)] | 0.0242 | 0.1819 | 1.0212 | 1.9267 | 2.2979 |
| $A_\lambda$[M06] | -0.0079 | -0.0043 | 0.0034 | 0.0118 | 0.0180 |
| $A_\lambda$[M06-2X] | -0.0043 | -0.0025 | 0.0050 | 0.0149 | 0.0245 |
| $A_\lambda$[M06-HF] | -0.0060 | -0.0050 | 0.0012 | 0.0104 | 0.0199 |

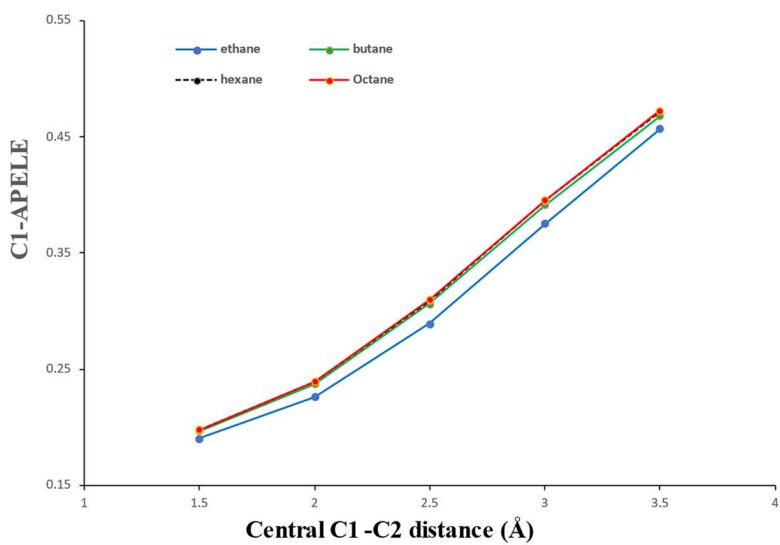

**Figure 1**. APELE of the central C atom in the alkanes as a function of the C-C distance.

**Table 3**: **Hexane**: Left-Right Correlation Analysis. The APELE values are per C atom of the central C-C bond.

| Diagnostic | 1.5Å | 2.0Å | 2.5Å | 3.0Å | 3.5Å |
|---|---|---|---|---|---|
| APELE | 0.1978 | 0.2391 | 0.3089 | 0.3950 | 0.4722 |
| $T_1$ | 0.0043 | 0.0068 | 0.0152 | 0.0150 | 0.0188 |
| $D_1$ | 0.0114 | 0.0355 | 0.0941 | 0.0847 | 0.1096 |
| %TAE[(T)] | 0.0567 | 0.1773 | 0.7456 | 1.0377 | 1.7525 |
| $A_\lambda$[M06] | -0.0077 | -0.0053 | -0.0001 | 0.0054 | 0.0095 |
| $A_\lambda$[M06-2X] | -0.0046 | -0.0036 | 0.0013 | 0.0077 | 0.0141 |
| $A_\lambda$[M06-HF] | -0.0058 | -0.0056 | -0.0018 | 0.0042 | 0.0105 |

Regarding the $A_\lambda$ method , the M06, M06-2X, and M06-HF functionals were used for the evaluation of this index since these functionals have a varying degree of Hartree-Fock exchange as: 27%, 54%, and 100% respectively. This may help to determine if the amount of Hartree-Fock exchange would significantly affect the diagnostic ability to detect NDC. The obtained $A_\lambda$ values all fall below the NDC threshold suggested by the authors of ref. [39], and misleadingly would indicate that no significant NDC is present in these molecules at any central C-C bond stretch. However, the $A_\lambda$ index does display qualitatively a proper trend with respect to the bond stretching. Similar to the other coupled cluster based diagnostics, the $A_\lambda$ diagnostic is not size-consistent and reflects the situation is a system averaged way (see Eq.(12) ).

Table 4: **Octane**: Left-Right correlation Analysis. The APELE values are per C atom of the central C-C bond.

| Diagnostic | 1.5Å | 2.0Å | 2.5Å | 3.0Å | 3.5Å |
|---|---|---|---|---|---|
| APELE | 0.1979 | 0.2394 | 0.3098 | 0.3956 | 0.4725 |
| $T_1$ | 0.0044 | 0.0063 | 0.0131 | 0.0132 | 0.0151 |
| $D_1$ | 0.0120 | 0.0353 | 0.0911 | 0.0840 | 0.0970 |
| %TAE[(T)] | 0.0743 | 0.1692 | 0.5989 | 0.8045 | 1.2862 |
| $A_\lambda$[M06] | -0.0075 | -0.0057 | -0.0017 | 0.0025 | 0.0056 |
| $A_\lambda$[M06-2X] | -0.0047 | -0.0039 | -0.0003 | 0.0046 | 0.0094 |
| $A_\lambda$[M06-HF] | -0.0057 | -0.0054 | -0.0025 | 0.0020 | 0.0067 |

We have also calculate the gross number of ELE $\bar{N}_u$ for a few selected molecules from the W4-11 database [12] (Table 5) . It is interesting that the two different indexes, $\bar{N}_u$ vs. %TAE[(T)] agree well qualitatively when comparing molecules with the same stoichiometric composition.

**Table 5.** Calculated gross number of ELE $\bar{N}_u$ vs the %TAE[(T)] index of selected small molecules.

| Molecule | C2H2 | CH2C | c-HCOH | t-HCOH | c-HONO | t-HONO | H2C2O | H2CCO |
|---|---|---|---|---|---|---|---|---|
| $\bar{N}_u$ | 0.8003 | 0.6898 | 0.7272 | 0.7117 | 1.2895 | 1.3062 | 1.1754 | 1.1636 |
| %TAE[(T)] | 2.10 | 1.90 | 2.30 | 2.30 | 5.30 | 5.40 | 3.2 | 2.5 |

Table 5 lists four pairs of such molecules selected from the W4-11 database. For each pair, the gross number of ELE is larger whenever %TAE is larger. To keep in mind that in most singlet cases our gross $\bar{N}_u$ values have an upper limit of 2 (diradical molecules), while values of $\bar{N}_u$ below 0.15 are typical for singlet single-bonded molecules like H$_2$ at equilibrium.

To compare more fully the different diagnostics, a statistical correlation analysis was done based on the calculated results. Table 6 contains the statistical correlation between the methods with respect to the trend of NDC upon increasing the C-C bond length of ethane. There is a good agreement between the diagnostics here, with the lowest correlation being ≈ 85%. This is because all diagnostics more or less increase their values as the C-C bond stretches, reflecting the increase of NDC.

**Table 6**. Correlation (in %) between the diagnostics upon increasing the central C-C bond length in ethane.

| | APELE | $T_1$ | $D_1$ | TAE[(T)] | $A_\lambda$[M06] | $A_\lambda$[M06-2X] |
|---|---|---|---|---|---|---|
| $T_1$ | 93.983 | | | | | |
| $D_1$ | 92.489 | 99.777 | | | | |
| %TAE[(T)] | 99.527 | 91.496 | 89.762 | | | |
| $A_\lambda$[M06] | 99.268 | 92.901 | 91.256 | 99.784 | | |
| $A_\lambda$[M06-2X] | 98.839 | 90.214 | 88.457 | 99.833 | 99.747 | |
| $A_\lambda$[M06-HF] | 97.704 | 87.519 | 85.713 | 99.287 | 99.119 | 99.781 |

Similar trend is seen from the data in Table 7, where the correlation between the methods is estimated across all considered molecules at their respective equilibrium bond distance. An exception is $A_\lambda$[M06-2X] there, showing an opposite trend compared to the rest of the diagnostics. Considering the $A_\lambda$[M06-HF] diagnostic, it only partially follows the correct trend marked by the APELE values. While most of the remaining methods do well at equilibrium, the analysis at 3.5Å in Table 8 shows a lack of size-consistency by all the methods but APELE, with high negative correlations with respect to the correct trend.

**Table 7.** Correlation (in %) between the various diagnostics at the equilibrium C-C bond length across all molecules.

|  | APELE | $T_1$ | $D_1$ | TAE[(T)] | $A_\lambda$[M06] | $A_\lambda$[M06-2X] |
|---|---|---|---|---|---|---|
| $T_1$ | 97.214 |  |  |  |  |  |
| $D_1$ | 97.799 | 99.963 |  |  |  |  |
| %TAE[(T)] | 96.765 | 99.982 | 99.899 |  |  |  |
| $A_\lambda$[M06] | 99.084 | 99.477 | 99.703 | 99.269 |  |  |
| $A_\lambda$[M06-2X] | -98.802 | -99.661 | -99.836 | -99.489 | -99.980 |  |
| $A_\lambda$[M06-HF] | 44.864 | 64.030 | 62.158 | 65.461 | 55.943 | -57.567 |

**Table 8.** Correlation (in %) between the diagnostics at 3.5°A C-C bond length across all molecules.

|  | APELE | $T_1$ | $D_1$ | TAE[(T)] | $A_\lambda$[M06] | $A_\lambda$[M06-2X] |
|---|---|---|---|---|---|---|
| $T_1$ | -47.501 |  |  |  |  |  |
| $D_1$ | -3.890 | 89.755 |  |  |  |  |
| %TAE[(T)] | -99.322 | 44.655 | 0.734 |  |  |  |
| $A_\lambda$[M06] | -97.176 | 64.709 | 24.589 | 97.107 |  |  |
| $A_\lambda$[M06-2X] | -99.400 | 53.475 | 10.777 | 99.458 | 98.974 |  |
| $A_\lambda$[M06-HF] | -99.371 | 54.429 | 11.891 | 99.319 | 99.103 | 99.992 |

### 4.2.2. Strength of NDC in acetylene vs vinylidene.

To extend the comparison among the different diagnostics we consider next a pair of interesting molecules presented on Fig.2 [41]. They are rather different in their properties but have formally the same stoichiometric composition. This allows a comparison of the relative strength of NDC in both structures using the APELE values as well as the gross number of ELE index $\bar{N}_u$. For instance, the carbon atoms in acetylene have relatively stronger NDC than either of the carbons in vinylidene, while APELE of the hydrogen atoms vary only slightly. The gross number of ELE $\bar{N}_u$ for acetylene and vinylidene is 0.800 and 0.690 respectively, while the %TAE[(T)] values are 2.1 and 1.9. This confirms that $\bar{N}_u$ for relatively small systems correlates qualitatively with wave-function based diagnostics.

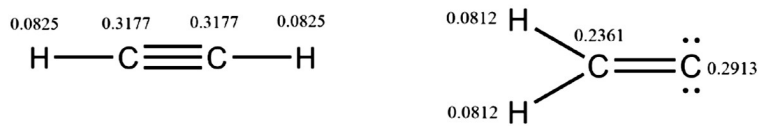

**Figure 2.** Comparison of APELE in acetylene (left) vs vinylidene molecules.

## 5. Other exemplary applications of the APELE method.

### 5.1. Study of bis-acridine dimers.

An interesting case study is the bis-acridine dimers, synthesized by Nakano et al.[20]. The diradical character of these compounds has been verified and is analyzed here using our diagnostics evaluated for the Model 8 and Model 12 structures presented in ref. [20] All calculations are done at B3LYP/6-31G* optimized geometries. The only difference between the two compounds is that Model 8 has a positive charge of 2+ while Model 12 is neutral. Both molecules have a singlet ground state. Model 8 has been shown to have strong diradical character and high second-hyperpolarizability corresponding to high TPA intensity, according to the experimental results [20]. Figure 3 shows our slightly simplified structure used for Model 8 (and similarly for Model 12) in the calculations. The obtained gross ELE values $\bar{N}_u$ are 12.834 and 12.281 for Model 8 and Model 12, respectively, indicating that Model 8 involves larger NDC than Model 12. However, in these systems size-consistency matters since the radical density formation is localized in certain sub-regions of the molecules. The spin-density distribution calculated in ref. [20] for Model 8 with spin-symmetry breaking (see their Figs.5 and 10) is mainly localized on the two N atoms and their nearest neighbors, carrying the bulk of the positive charge and radical charge altogether. Our APELE results are in good agreement with the findings of Nakano et al., even though we do not use spin-symmetry breaking. Table 9 contains the APELE of the atoms that are carrying most of the diradical formation: (HN1+C2+C3) on one side (the N atom with its nearest neighbors, see Fig.3) and its symmetric group counterpart (HN6+C7+C8) on the other side, each group carrying a total of 1.035 unpaired radical charge. We also find some noticeable accumulation of APELE around each of the two oxygen atoms (O4 +C5= 0.76 = O9+C10) (Table 9), each oxygen being situated on the symmetrically opposite side of the N atoms. This indicates a possibility for tetra-radical formation in this compound or at least that NDC has an increased significance in the vicinity of these oxygens too. Once more, the ability of probing the strength of NDC locally/regionally within a given system brings some more detailed insight.

| HN1, HN6 | C2, C8 | C3, C7 | O4,O9 | C5,C10 |
|---|---|---|---|---|
| 0.457 | 0.284 | 0.294 | 0.432 | 0.324 |
| Total: | **1.035** | | **0.756** | |

**Table 9**. Calculated APELE values of the atomic groups in Model 8 carrying most of the radical density.

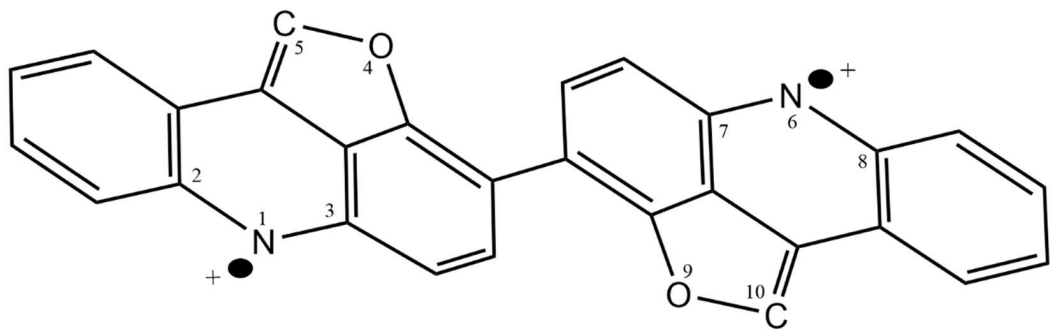

**Figure 3**. A slightly simplified structure of Model 8 used in the calculations.

### 5.2. Study of the model system PQM

Much attention has been devoted to a particular class of organic diradicals that show an unusually high nonlinear optical (NLO) response [17, 19, 22]. Molecular design of efficient NLO systems is a subject of intense research due to their importance for photonics and optoelectronics such as optical switches, 3D microfabrics, photodynamic therapy and other. Recent studies have shown that the degree of diradical character in systems like p-quinodimethane (PQM), is related to the large second-order hyperpolarizability ($\gamma$) observed in these systems [17, 20-23].

PQM (Fig.4) is a model system studied first by Nakano and collaborators using their y-factor index (see Eq.(8)) in the search for a connection between the degree of diradical character and the second hyperpolarizability [17]. Our APELE analysis allows us to reveal from a somewhat different angle the strongly correlated nature of the NLO active diradicaloids. The p-quinodimetane (PQM) is an interesting example itself of a normal singlet state at equilibrium that transforms into an open-shell singlet upon stretching of certain characteristic bonds [17]. First we optimized the geometry of the PQM with B3LYP following the recommendation of ref.[17]. The lowest-energy geometry obtained with B3LYP has a typical alternating C-C bond pattern in the ring, similar to the one reported in ref. [17]. Bond length alternation in carbon ring systems is a known indication of $\pi$-electrons localization on the individual C-C bonds. This localization is sustained by strong ND left-right correlation. Similar effects have been observed in other polycarbon systems [42], and in periodic systems [43]. In PQM this effect is accompanied with loss of aromaticity, which is observed also in other carbon ring systems with alternating C-C bond lengths [42]. Our results presented in Table 10 show that the PQM di-radical charge estimated by APELE is partly delocalized around the C manifold, with an emphasis on the two terminal C-CH$_2$ group of atoms. It gradually increases with the symmetric stretch of the C-CH$_2$ bond distance on both sides of the ring. More pronounced is the change of the group-APELE $Fr$(C-CH$_2$), which shows that the increased NDC upon bond stretching is mainly confined locally within these group of atoms where the bond stretching occurs. The global NDC index $\bar{N}_u$ is relatively less sensitive to the stretching, which shows again the importance of examining the effect of NDC locally.

Considering the diradical formation upon bond-stretching, a few DFT indexes have been proposed so far based on UKS calculations [17, 22, 23]. The $y$ index was mentioned already (Eq. (8)). A difficulty for this approach is the fact that not always the UKS or UHF calculations lead to a lower energy state with broken symmetry and localized radical spins. For example, we performed a stability analysis of the RKS vs UKS state of the PQM using RB3LYP and UB3LYP respectively. At the equilibrium PQM geometry, there is no RKS-UKS instability of any kind, and the ground state energy remains RKS. Our APELE analysis on the other hand shows that there is a good amount of NDC as evidenced by the $\bar{N}_u$ value of 2.78. In this case one can hardly claim a diradical formation based on the UKS treatment alone, even though the $y$ parameter value is also nonzero (Table 10). True RKS-UKS instability appears only when both C-CH$_2$ bond lengths are stretched to about 1.6 Å or so and longer.

| R$_1$, R$_1$ | R$_2$ | R$_3$ | $F_r$(CCH$_2$) | $F_r$(C$_1$) | $F_r$(C$_2$) | $F_r$(C$_9$H) | $\bar{N}_u$ | $Q_r$ | $y^c$ |
|---|---|---|---|---|---|---|---|---|---|
| 1.348 (eq) (1.351 $^a$) | 1.457 (eq) (1.460) | 1.343 (eq) (1.346) | 0.715 [0.713$^b$] | 0.253 [0.255$^b$] | 0.303 [0.301$^b$] | 0.338 [0.339$^b$] | 2.78 | 0.514 | 0.146 |
| 1.4 | 1.4 | 1.4 | 0.767 | 0.250 | 0.348 | 0.269 | 2.92 | 0.525 | 0.335 |
| 1.5 | 1.4 | 1.4 | 0.854 | 0.264 | 0.390 | 0.346 | 3.06 | 0.558 | 0.491 |
| 1.6 | 1.4 | 1.4 | 0.905 | 0.278 | 0.435 | 0.349 | 3.21 | 0.564 | 0.626 |
| 1.7 | 1.4 | 1.4 | 0.981 | 0.294 | 0.482 | 0.349 | 3.36 | 0.584 | 0.731 |
| 1.8 | 1.4 | 1.4 | 1.057 | 0.309 | 0.530 | 0.350 | 3.52 | 0.601 | |
| 2.0 | 1.4 | 1.4 | 1.206 | 0.342 | 0.621 | 0.350 | 3.81 | 0.633 | |

**Table 10.** APELE of atoms and group of atoms in the PQM diradicaloid ($F_r$(A)). $^a$ Equilibrium geometry estimates of ref. [17]; $^b$ Values obtained perturbatively based on RB3LYP SCF solution. $^c$ The $y$ values are from ref.[17]. The series of PQM geometries follow the pattern given in ref.[17]: R$_1$, R$_2$, R$_3$ are the three key bond distances in PQM, one of which, R$_1$, is gradually stretched in a symmetric fashion (from both sides of the ring, See Fig.4).

Another group-APELE index in the case of PQM that shows the relative degree of concentration of radical charge on the terminal C-CH$_2$ groups is $Q_r = 2(F_r(\text{CCH}_2) / \bar{N}_u$, where $F_r$(CCH$_2$) is the group-APELE in the union of these four atomic regions as mentioned above, Fig.4. These are the atoms of the C-CH$_2$ fragment where most of the APELE charge is located. At equilibrium geometry our estimate of $Q_r$ here is about 0.514. It gradually increases to about 0.633 at stretched C-CH$_2$ distance of 2.0 Å. It is worth noting that the total number of radicalized electrons $\bar{N}_u$ is larger than 2 even at equilibrium. This is because there is some small odd-electron population also on other C atoms from the ring, besides the di-radical formation (Fig.4). This is mainly due to the above-mentioned localization of the ring $\pi$ electrons. Those findings are not accessible from the values of the $y$ index alone presented in Table 10, which is a global characteristic of the system.

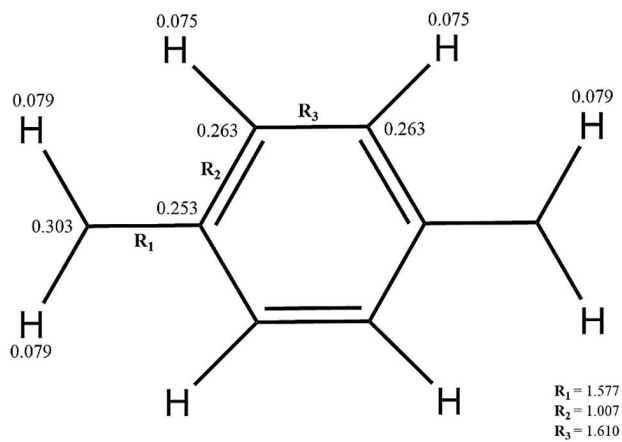

**Figure 4.** The PQM structure optimized with B3LYP using G3LARGE basis set and (128,302) unpruned grid. APELE values are in red, calculated bond order values are in green.

### 5.3. Study of the ethylene addition to Ni dithiolene reactions

Metal bis(dithiolene) complexes have been widely studied because of their unusual chemical, redox, and optical properties related to the specific nature of the dithiolene ligand [44] [45] [46] [47]. Their complex character makes it difficult to assign unambiguously the oxidation state of the metal center, as pointed out in the studies of Dang et al.[48] on the transition states and intermediates of the ethylene addition to Ni(edt)$_2$ (edt=S$_2$C$_2$H$_2$).
They examined the accuracy of several DFT functionals in comparison with CCSD, CCSD(T), CASSCF and CASSCF-PT2 estimates. To inquire about the oxidation state of the metal center, we use the APELE approach to estimate the population of effectively localized electrons in the '1H' and '2H' reaction species in the nomenclature of ref.[48] (Figs.5-7). Large APELE values mean the corresponding atoms are less involved in the present covalent bonding but remain highly reactive due to the trend of their unpaired radical spin to saturate. The comparison between the $\bar{N}_u$ and $T_1$ diagnostic values in this case shows that they correlate relatively well (Fig.5).

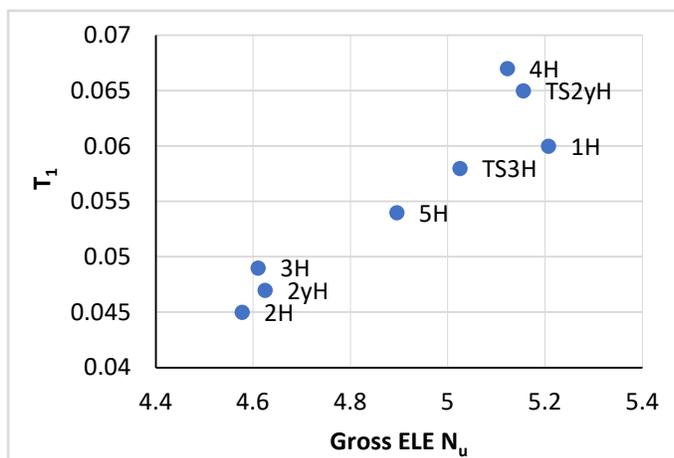

**Figure 5.** Correlation between $\bar{N}_u$ and $T_1$ for the different reaction species.

The variations among different reaction species can be further analyzed based on $\bar{N}_u$ and the individual or group APELE values, which are not available from the $T_1$ analysis alone.

We have performed also relative energy analysis of the ethylene addition to Ni dithiolene using the KP16/B13 functional [5]. The calculated relative energies between the different reaction species are compared with the CCSD(T) results of ref.[48]. Figure 6 shows the correlation between these two sets of results. Our predicted relative energies are somewhat higher than CCSD(T), but still, they correlate reasonably well with the CCSD(T) estimates.

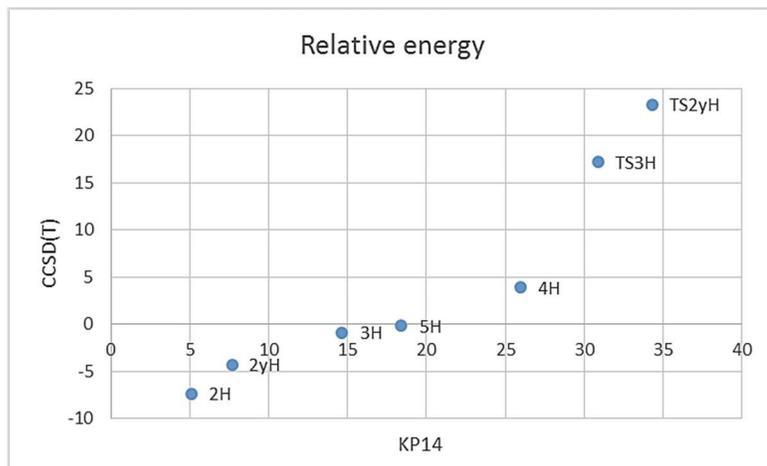

**Figure 6.** Correlation between the relative energy estimates with KP16/B13 and CCSD(T).

To obtain a better impression on the performance of contemporary functionals vis-a-vi CCSD(T), we ran a linear regression of the relative energy values reported with M06, HSE06, and wB97XD in ref. [48] and compared them with our KP16/B13 results. The statistics are summarized in Table 11. KP16/B13 has a larger intercept, but the slope of its correlation with CCSD(T) is closer to 1, and the correlation index is good too.

|             | KP16/B13 | M06  | HSE06 | wB97XD |
|-------------|----------|------|-------|--------|
| Intercept   | 15.4     | 6.2  | 2.0   | 8.8    |
| Slope       | 0.92     | 1.15 | 1.1   | 1.75   |
| Correlation | 0.88     | 0.83 | 0.88  | 0.91   |

**Table 11.** Correlation of the results from several functionals with the CCSD(T) benchmark for the relative energies in the reaction of ethylene addition to Ni(edt)$_2$.

The results from the APELE analysis of some of the reaction species (Fig.7) are summarized in Table 12. We observe a large decrease of APELE on the two upper S atoms when going from 1H to 2H reaction species. This is due to the bonding of the carbon atoms of ethylene to these two S atoms in the 2H product. These new S-C bonds cause some decrease of APELE on the Ni center, which means that a fraction of the odd electrons on Ni in 1H is now engaged in enforcing the Ni-S bonds in 2H. Therefore, our APELE analysis predicts some increase of the oxidation state of the Ni center when going from the initial 1H structure to the 2H product.

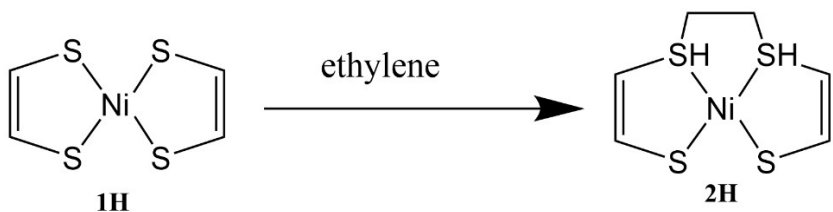

**Figure 7.** 1H and 2H reaction species: 1H is the initial reactant, 2H is one of the two possible products per ref.[48].

| Atoms | APELE 1H | APELE 2H | 2H -1H |
|---|---|---|---|
| Ni | 0.666 | 0.563 | -0.104 |
| S3 | 0.503 | 0.374 | -0.129 |
| S2 | 0.503 | 0.463 | -0.040 |
| S1 | 0.503 | 0.374 | -0.129 |
| S4 | 0.503 | 0.462 | -0.041 |
| C2 | 0.338 | 0.315 | -0.023 |
| C3 | 0.338 | 0.298 | -0.040 |
| C4 | 0.338 | 0.314 | -0.024 |
| C1 | 0.338 | 0.300 | -0.039 |
| C | 0.242 | 0.207 | -0.035 |
| C | 0.242 | 0.211 | -0.031 |

**Table 12.** Calculated APELE values for the atoms involved in the reaction of ethylene addition to $Ni(edt)_2$.

### 5.4. Study of Multireference Character of Transition-Metal Compounds

A variety of materials contain transition metal (TM) elements as essential ingredients. Those compounds often have strong multireference character mainly due to the partly filled inner d-shells of the TM centers. The APELE method could be advantageous for studying such compounds, especially those involving large ligands. We explore this possibility using the ccCA-TM-11 (correlation-consistent composite approximation for transition-metals 2011) database [49]. It contains 193 complexes involving first row (3d) transition metals, ranging from the monohydrides to larger organometallics such as $Sc(C_5H_5)_3$ and clusters such as $(CrO_3)_3$. The heat of formation for each complex has already been reported in the literature. The multireference character of those TM complexes were also studied by the same authors in a subsequent paper using $T_1$, %TAE, $D_1$ indexes, and $C_0$ of CASSCF [36]. Table 13 lists the APELE on the TM centers Ni and Cr in a series of exemplary carbonyl complexes. All these molecules there are singlets. The APLELE on the TM atoms reflect the rather localized nature of its valence d-like electrons. The %TAE and $T_1$ results[36] show a decrease trend when more ligands are added to Ni, which is due to the size inconsistency of these diagnostics. APELE, on the other hand, shows generally an increase trend for the Ni series.

| Molecule | APELE | %TAE | $T_1$ |
|---|---|---|---|
| NiCO | 0.63 | 11.1 | 0.046 |
| $Ni(CO)_2$ | 0.67 | 7.5 | 0.037 |
| $Ni(CO)_3$ | 0.66 | 6.4 | 0.031 |
| $Ni(CO)_4$ | 0.69 | 6.1 | 0.031 |
| $Cr(CO)_3$ | 0.97 | 6.5 | 0.054 |
| $Cr(CO)_4$ | 0.96 | 5.8 | 0.036 |
| $Cr(CO)_5$ | 0.95 | 5.5 | 0.032 |
| $Cr(CO)_6$ | 0.94 | 5.3 | 0.028 |

**Table 13.** Calculated APELE values on Ni and Cr in a series of carbonyl complexes.

Interestingly, both the CCSD(T)-based estimates and the APELE values decrease in the Cr series, when number of the ligands increases, albeit the APELE values decrease relatively slower. As far as the APELE method concern, this behavior is not quite clear yet and requires some further investigation.

## 6. Conclusions

Nondynamic correlation is important in many cases, but it is expensive and hard to take into account . Most approximate DFT functionals fail to describe properly this correlation. A number of diagnostics have been proposed that allow one to estimate the quality of a given single-reference solution in this respect. Most of these diagnostics are, however, not size-consistent, while remaining computationally expensive. In this work we use the DFT method of atomic population of effectively localized electrons and the gross number of effectively localized electrons as novel diagnostics in this vein. These are compared with several existing diagnostics of nondynamic correlation on selected exemplary systems. The APELE method is generally in good agreement with the currently used diagnostics, while having the advantage of being size-consistent, less costly and does not involve unrestricted spin-symmetry breaking. It becomes particularly informative in cases involving bond stretching or bond breaking as in the series of linear alkane chains considered here. The APELE method is used also to explore some specifics of organic diradicaloids like the bis-acridine dimer and the p-quinodimethane having unusually high nonlinear optical response. The calculated APELE values point out on nonnegligible nondynamic correlation in these systems that is localized within certain specific subdomains of the molecules. We tested the method also on the complicated reaction of the ethylene addition to Ni dithiolene, where our results shed some light on how the oxidation state of the Ni center may change when going from the initial reactant (1H) to the product (2H).

**Acknowledgement**
JK thanks Dr. Pachter for very helpful discussions on the topic of this paper. The authors thank Dr. Fenglai Liu, Mathew Wang and Dwyane John for assistance. This work received support from Air Force Research Laboratory of U.S. Department of Defense under the AFRL Minority Leaders - Research Collaboration Program, contract FA8650-13-C-5800 (Clearance Authority: 88ABW-2016-5818), and from the National Science Foundation (Grant No. 1665344). EP thanks Dennis Salahub for the long years of mentorship, friendship and support.